# High entropy ceramics for applications in extreme environments

T.Z. Ward[1], R.P. Wilkerson[2], B.L. Musicó[2], A. Foley[3], M. Brahlek[1], W.J. Weber[4], K.E. Sickafus[3], A.R. Mazza[3,*]

[1]Materials Science and Technology Division, Oak Ridge National Laboratory, Oak Ridge, Tennessee 37831, USA

[2]Sigma Division, Los Alamos National Laboratory, Los Alamos, New Mexico 87545, USA

[3]Materials Science and Technology Division, Los Alamos National Laboratory, Los Alamos, New Mexico 87545, USA

[4]Department of Materials Science & Engineering, University of Tennessee, Knoxville, TN, 37996, USA

[*]armazza@lanl.gov

**Abstract**

Compositionally complex materials have demonstrated extraordinary promise for structural robustness in extreme environments. Of these, the most commonly thought of are high entropy alloys, where chemical complexity grants uncommon combinations of hardness, ductility, and thermal resilience. In contrast to these metal-metal bonded systems, the addition of ionic and covalent bonding has led to the discovery of high entropy ceramics. These materials also possess outstanding structural, thermal, and chemical robustness but with a far greater variety of functional properties which enable access to continuously controllable magnetic, electronic, and optical phenomena. In this perspective, we outline the potential for high entropy ceramics in functional applications under extreme environments, where intrinsic stability may provide a new path toward inherently hardened device design. Current works on high entropy carbides, actinide bearing ceramics, and high entropy oxides are reviewed in the areas of radiation, high temperature, and corrosion tolerance where the role of local disorder is shown to create pathways toward self-healing and structural robustness. In this context, new strategies for creating future electronic, magnetic, and optical devices to be operated in harsh environments are outlined.

1. **Introduction**

Many existing and future technologies require materials that can withstand and function reliably under extreme environments involving high temperatures, pressures, chemical surroundings, ionizing radiation, or external fields. These varied environments each come with specific types of damage that can accumulate and evolve with continued exposure. As examples, metal alloys can succumb to enthalpic effects, dealloying, and phase segregation through thermal and pressure cycling (Fig. 1a) [1], while ionically and covalently bonded systems can have additional complications arising from the anion sublattice where bond breaking can lead to vacancy formations [2], morphotropic structural phase transformations [3], or structural decoherence (Fig. 1b). Understanding and countering degradations such as these is an ongoing challenge to ensure long lived functional materials that can operate in a variety of extremes.

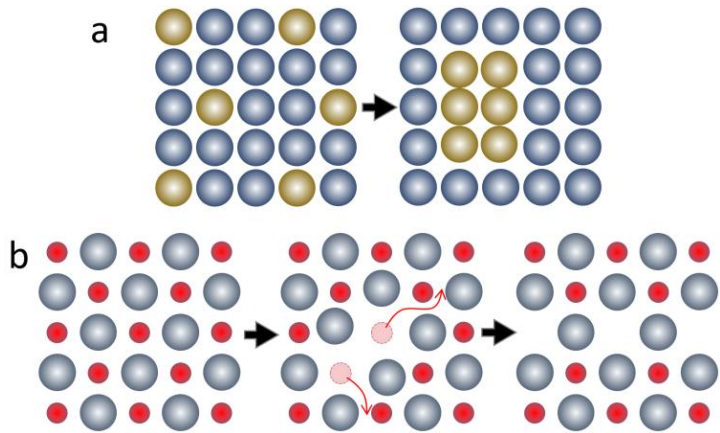

Fig. 1. Examples of materials evolutions under extreme temperatures. a) High temperatures in alloys can lead to phase segregation and clustering of dissimilar environments, while b) Ceramics are susceptible to anion migration leading to vacancies, phase transformation, and decoherence.

Historically, materials development has moved from simple low complexity materials in the iron age, to steel in the industrial age, and into superalloys and multicomponent materials with highly tailored compositions in the modern day [4]. Properly tailored disorder has been key for designing materials with increased robustness compared to perfect monoatomic crystalline materials especially when designing for harsh environments. Recent advances provide new opportunities to understand and control disordered structures across multiple length scales to further improve material stability [5–9].

In this perspective, we consider high entropy ceramics that may provide new access to functional applications in extreme environments. The paradigm of "high entropy" materials began with the development of high entropy alloys [10] in 2004 [11]. "High entropy" was used to describe the stabilization of ideally mixed multi-component materials by minimization of the Gibbs free energy by a large configurational entropy [10,12,13]. The advent of high entropy ceramics [13] (HECs), having an added degree of stabilization by the anion sublattice, a much broader range of crystal structures, and strongly correlated functional (magnetic, electronic, etc.) behaviors did not appear until 2015 with the invention of high entropy oxides [14]. A discussion of the definition of "high entropy" can be found elsewhere [12], though we point out in this work we use the broad definition of an ideally mixed solid state solution containing 5 or more elements on like sublattice sites. In these materials, even single crystals can provide lattice distortions from differences in atomic size that can lead to solid solution strengthening, with precise control over hardness and ductility [15]. High-symmetry crystal structures are entropically favored and have improving thermal stability. The ability to include many different cations on these stable lattices provide access to functional design strategies that were not previously possible. The influence of compositional disorder on entropic stabilization also acts on the larger lattice to help restrict grain growth, improve mechanical properties, and greatly reduce thermal conductivity through increased phonon scattering [16,17]. Additionally, these materials have been shown to have excellent corrosion resistance [18]. These structural benefits are well explored in high entropy metal alloys, but the recent advance of adding an anion sublattice is showing that inter-atomic complexity can also be used to generate magnetic, electronic, optical, and ferroic properties that arise from frustrated nearest neighbor interactions [19–21].

There is a critical need to develop new materials to endure such environments without failure [22]. Advances in characterization tools and modeling techniques are allowing unprecedented access to the fundamental atomic and molecular mechanisms driving materials stability across broad time and length scales. A central and broadly observed truth in many of these studies is that disorder introduced by defects and microstructural inhomogeneities provides mechanisms for impeding deformation and damage accumulation in materials. At the most local scale, point defects such as vacancies and interstitials can immobilize dislocations, slowing deformation processes. These inhomogeneities can also impact atomic diffusion rates, which can promote recovery/prevention of damage during exposure to thermal and structural shock. Additionally, the regions around defect sites can act as sinks for impurities and transmutation products which can reduce their influence on macroscopic properties through localization and trapping. Moving up in scale, the interfaces between adjacent grains or crystallites in a polycrystalline material provide barriers to dislocation motion. These grain boundaries can act to restrict slip and plastic deformation, resulting in improved strength and ductility. The high density of defects at grain boundaries also makes them preferred sites for trapping impurities and accommodating strains, which can further improve stability. Relying on the above mechanisms for structural stability potentially results in the cost of inconsistency in functional stability, as they rely on extrinsic structural evolutions that necessarily influence the local electronic, magnetic, and optical properties of the underlying crystal lattice.

In summary, introducing multiple cations enables the entropy-driven stabilization of new compositions with tailored properties from lattice disorder and distortions. These are enabled by a minimization of the Gibbs free energy of formation by maximizing the configurational entropy in the system. This facilitates the design of materials which simultaneously combine features like high hardness and fracture toughness, or low thermal conductivity and high elastic modulus with new levels of functional control related to ion transport [23], dielectric response [24], magnetic behavior [19], and catalytic reactivity [25]. While this perspective is focused on experimental work, it is imperative to include the work being done to model and guide synthesis of HECs. We include here works on materials discovery [26,27], stability [28], and more general reviews covering theoretical work [13,29]. We note much of the early theoretical work was focused on high entropy carbides, guiding much of the work outlined below, but has expanded to include (for example) oxides [26,30], nitrides [31], and borides [32]. This review is separated into two main sections. The first, "Structural Robustness", focuses on the properties of HECs which cater to their improved behavior in chemical, temperature, and radiation extremes. The second, Functional High Entropy Ceramics in Extreme Environments" focuses on the emergent electronic, magnetic, and optical properties of HECs and how they might cater to applications such as microelectronics and sensing in extremes.

2. <u>Structural Robustness</u>

**2.1 The advantages of chemical complexity**

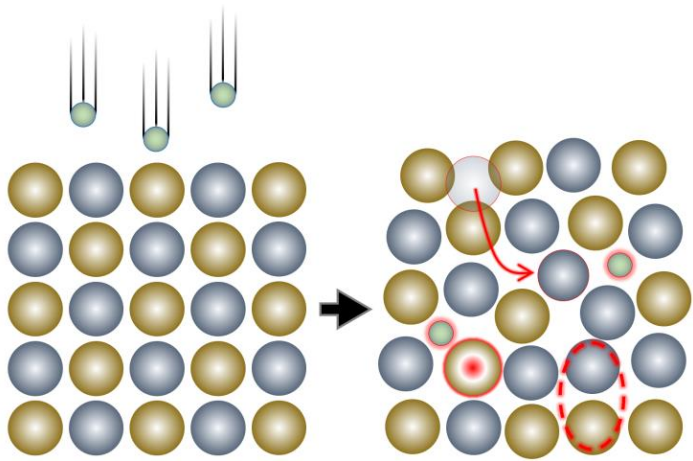

Fig. 2. Exposure to reactive and energetic particles (left) can drive multiple defect types including vacancies, interstitials, site reversals, activation, and implantation (right).

Complexity offers a large design space in terms of materials discovery – where the added configurational entropy lends towards a favorable Gibbs Free energy of formation in a wide array of compositions. Chemical disorder enables a different crystallographic space and a matrix for entrapment that can improve phase stability in extreme environments. The resistance to structural perturbation makes this class of materials interesting in the exploration for new materials with properties such as high ionizing radiation resistance, extreme temperatures, and chemical barrier coatings. Ion irradiation and other radiation effects introduce atomic disorder to materials, destroying crystallinity, creating strain, and forming vacancies and dislocations, as shown in Fig. 2. HECs, which minimize the Gibbs free energy of formation by increasing the number of cations or anions occupying like lattice sites, may demonstrate a resilience to irradiation induced damage. The atomic structures of multicomponent high entropy materials are complex – potentially hosting a range of microstructures, local atomic coordination, valences, chemical potentials, and beyond. Because high entropy materials can be comprised of numerous cations with a range of electronegativity and size, they exhibit local atomic structures which bend conventional understanding of bonding in ionic solids. For example, they can exhibit an inherent, random distribution of bond angles and host complex multivalences that enable the accommodation of intersite defects and phase frustration [9,19,33,34]. Considering the ability to stabilize materials beyond known phase diagrams provided by the high entropy approach, more subtle anomalies with contradictions to accepted principles regarding ionic compound formation are possible.

As an example, Pauling's 2nd rule, known as the *electrostatic valence principle*, argues that the valence of a central anion is neutralized by first nearest neighbor (n.n.) cations. When considering HECs, the neutralization of charge becomes quite an interesting conundrum – with a vast landscape of potential cation valences and local chemistries. This principle is generally satisfied (for instance in $MgAl_2O_4$ spinel) but there are violations. In fact, ordinary (binary) pyrochlores do not satisfy Pauling's 2nd rule, in the sense that individual anions in the pyrochlore lattice are either over-bonded or under-bonded by n.n. cations. In ionic solids, the usual crystal chemical response to over-bonding or under-bonding is bond length change. Over-bonding leads to shorter bonds, under-bonding leads to longer bonds. The shorter bonds have greater bond strengths, the longer bonds have diminished bond strengths. The relationship of over-bonding/under-bonding to bond length/bond strength is known as the Zachariasen effect [35].

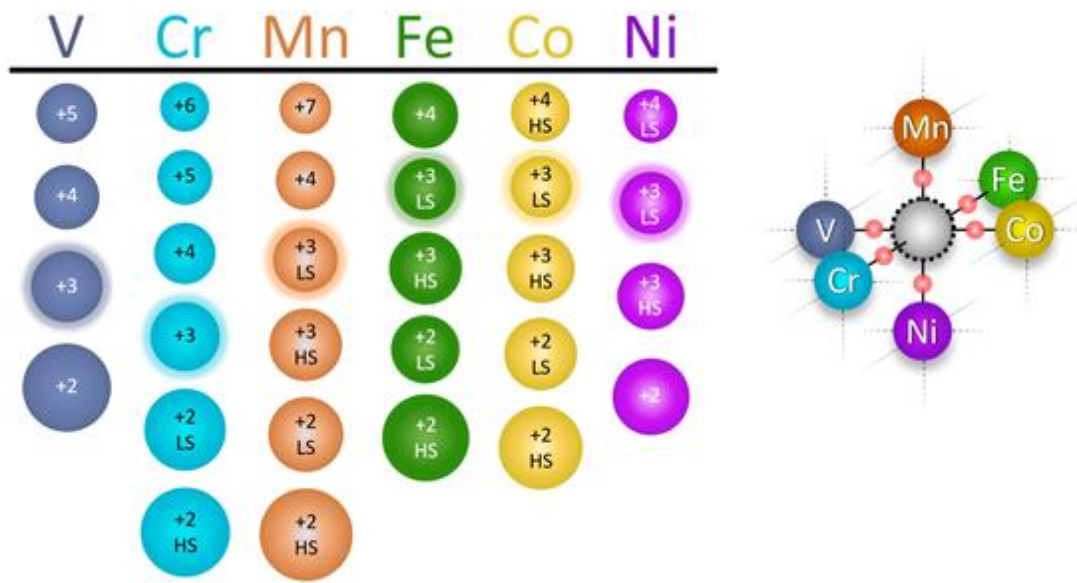

Fig. 3 The range of possible valences for 3d cations from vanadium to nickel. The cations are scaled based upcon the Shannon and Prewitt ionic radii. In the right panel the concept of partial charge transfer is represented for a cation with many different nearest neighbors, each with unique electronegativity, preferred charge, and size impacting the equilibrated valence.

Here, lattice distortion is generated from coordination polyhedra deviating from ideality to become irregular polyhedral. In HECs, large deviations from ideality have been observed in perovskites [9,33], fluorites [36], and rocksalts [37] and is expected to exist in a number of other systems. Cation coordination polyhedra around central O anions have been shown to be severely distorted due to the cations at the vertices of any polyhedron being comprised of dissimilar cationic species.

It is also interesting to note that these bond length/bond strength concepts are intricately linked to electronegativity. Considering the complex landscape of chemical environments in HECs, the local structure should be highly impacted by this connection. The Sanderson principle of electronegativity equalization links bond strength, bond length, and electronegativity [38,39]. This principle states that in forming a cation/anion bond, partial charge is transferred from the cation (low electronegativity) to the anion (high electronegativity), until the electronegativities of the partially charged ions become equal. At this stage of becoming equal, charge transfer ceases. The extent of charge transfer (magnitude of d) determines the relative ionicity/covalency of the interatomic bond (i.e. large d, more ionic; small d, more covalent). Interestingly and to this point, in high entropy materials the outcome of charge transfer can be quite nontrivial. Taking the example of 3d metals, a wide array of possible valences and spins are possible at the local level when randomly populating a cation site. Depending on the electronegativity and preferred charge, these sites can share partial charge or even compensate on the global scale. This is illustrated in Fig. 3. The smaller the partial charge, the higher the intrabond electron density and therefore the shorter/more covalent the bond. Reciprocally, the larger the partial charge, the lower the intrabond electron density and the longer/more ionic the bond. To put all of this in the perspective of high entropy materials, it is anticipated that because the electronegativity of each cation is different, bond lengths and bond strengths, as well as relative ionicity/covalency, will vary between each bond, even within a single coordination polyhedron. Consequently, coordination polyhedra will be highly irregular, and local crystal structure may altogether change, because the principles of crystal formation no longer obey ionic structure rules. There have already been studies suggesting such a complex landscape of local charge [21,33,34,40–43] with direct effects on local structures, indicating these complex microstructures are present in high entropy materials. The resulting severe lattice strains in high entropy materials should lead to limited atom mobility, resulting in, for example, decreased radiation-induced structural degradation due to the favoring of interstitial-vacancy point defect recombination. Below, we explore the current state of the art in the use of HECs in extreme environments, focusing first on their performance in thermal and chemical extremes required of barrier coatings.

**2.2 Barrier Coatings**

One of the more promising applications for HEC materials is in the fields of thermal and environmental barrier coatings. There is quite a bit of overlap in the needs of these applications as the functional behaviors of candidate materials must persist in extreme environments. Examples of these environments can be broad but are generally hazardous thermal and chemical environments such as hypersonic vehicle control surfaces, the interior of aircraft gas turbine engines, and light water reactor (LWR) nuclear fuel cladding. Generally, properties of foremost interest for barrier coatings are thermal stability, low thermal conductivity, chemical resistance, and substrate compatibility (adhesion and expansion) [44,45]. It becomes immediately apparent that HECs present as an attractive option, with intrinsically low thermal conductivity [16,46] and improved thermal stability [37,47], with other properties of interest likely being tailorable via structure and/or composition.

Taking a few examples of extreme environments utilizing barrier coatings, in gas turbine engines coating interactions with calcia-magnesia-alumina-silicate (CMAS) glass are a major issue. CMAS is formed from engine ingestion of volcanic ash, sand, and runway dust [18,48], and can corrode oxide coatings and exacerbate spallation of coatings off the base material. As such there has been an explosion of research into the area of HECs for use in barrier coating applications with many promising candidates and crystal structures. The current state of the art for barrier coatings is yttria doped zirconia (YSZ) in a fluorite structure. YSZ is susceptible to phase transformations when used at temperatures in excess of 1200°C [49–52]. High entropy fluorite structures are the logical advancement of barrier coatings into high entropy space. Compositions such as $(ZrCeHfY)_{0.25}O_{1.8-1.9}$ as well as $(ZrCeHfYRE)_{0.2}O_{1.8-1.9}$ (RE = La, Nd, Sm) have shown reduced thermal conductivity and improved hardness, with early evidence of durability up to 1200°C [53]. Aluminum has also been used to dope the similar composition of $(ZrCeHfYAl)_{0.2}O_{1.8}$ where the oxygen stoichiometry is impacted by partial valence reduction of Ce, Zr, and Hf, favorably impacting the thermal expansion coefficient during heating and cooling while still maintaining the fluorite structure [54].

Towards the same end, zirconate structures such as $La_2Zr_2O_7$ are an alternate to YSZ as a barrier coating material, with increased phonon scattering due to heavier rare earth elements and higher oxygen concentrations. Zirconates also do not exhibit high temperature transformations in the solid phase and have a high melting point, aiding in their ability to act as barrier coatings [55]. The $A_2B_2O_7$ formulation for these materials can form in a disordered fluorite structure or a pyrochlore-type, with $A^{3+}$ and $B^{4+}$ cation sublattices, though it is possible to include off-nominal charged cations into these sites [56]. This family of compositions presents even more possible combinations with entropic disorder in the A-site [50,51,56–62], B-site [52,63,64], or both [48,49,56,65,66], allowing for a massive compositional space for tuning of properties of interest. Of these, A-site disordered zirconates have received the most attention, with zirconium, hafnium, or cerium being the single B-site occupant. These show great promise to exceed the already extraordinarily high melting points of the individual parent oxides, which directly impacts the high temperature stability and functionality of future HEC coatings of this type.

In addition to these two families of structures for these coatings, silicate [18,67,68] and disilicate [68–71] HECs have also begun to emerge as promising barrier coating materials. Silicon containing ceramics as barrier coatings have distinct advantages such as compatibility with other ceramics such as SiC and prevention of oxidation of underlying layers. One example of a high entropy monosilicates is $(Er_{0.2}Tm_{0.2}Yb_{0.2}Dy_{0.2}Y_{0.2})_2SiO_5$ [67], which exhibits improved corrosion resistance as compared to binary silicates to 1300°C. Disilicates have shown to perform better as barrier coating materials in regards to their improved thermal expansion mismatch and hydrothermal corrosion resistance [68,69]. As the field is growing, a handful of other materials such as aluminates [72,73] and phosphates [74] have been pursued in the HEC space for coatings, demonstrating the room to mature and explore the infinite palette of compositions for use in different extreme environments.

**2.3 Ion irradiation and radiation hardness**

Studies exploring the resilience of high entropy ceramics to radiation effects have begun to emerge in the literature. This research can be broken into three directions, which are the focus of the following sections: irradiation of high entropy carbides, self-irradiation of actinide containing high entropy ceramics, and irradiation effects of high entropy oxides.

*2.3.1 High Entropy Carbides*

Carbides, owing to their very high melting points and mechanical properties, have long been a primary focus of materials research for extreme environments. The high entropy analogues to simple AC binary carbide materials (*A* = Mo, Nb, Ta, V, W, for example) have been a promising area of research and provide some of the first evidence for physical properties which extend beyond the performance of the parent compounds [15,75,76]. The continuously tunable combination of elements and tunable properties of high entropy carbides make them uniquely interesting in the exploration of radiation hard materials, particularly in high temperature environments like those in future fusion reactors, advanced nuclear reactors, and other nuclear materials applications. A key necessity in this area of research is irradiation tolerance, which has been explored in a series of carbides inspired by the higher tolerance of high entropy alloys as compared to traditional alloys. The exploration of radiation hardness of high entropy carbides has been limited but quite promising [47,77].

One early example is a self-irradiation (1 MeV carbon ion) study on $(MoNbTaVW)_{0.2}C$. This work revealed a small, saturated expansion of the lattice due most likely to interstitial carbon of ~0.6% at room temperature, with a reduction in this expansion for elevated temperatures [78]. Using transmission electron microscopy (TEM), Zhu et. al demonstrate that the effects of the irradiation diminish at higher temperature and that no amorphization was observed at any of the implantation conditions. The effect of irradiation on the structure peaked at a fluence of $~10^{16}$ ions/cm$^2$ at room temperature but were recoverable at higher fluence and temperatures owing to the mobility of anions (carbon in this case) at cations repairing local defect networks.

Heavy ion irradiation of high entropy carbides has also been explored in the $(ZrTaNbTi)_{0.25}C$ system. These samples were irradiated by 3 MeV $Zr^{2+}$ ions, at 25, 300, and 500°C, and the resulting change in structure was analyzed using x-ray diffraction and TEM [79]. Remarkably to the highest dose in the study, an equivalent dose of 20 dpa, the structure of the high entropy carbide remained stable with a maximum expansion of 0.2%. Furthermore, no segregation or void formation was observed in the bulk or at grain boundaries as a result of implantation.

The light atom (He) irradiation tolerance of a similar high entropy carbide, $(ZrTiNbTaW)_{0.2}C$, was also studied through implantation with 540 keV ions up to a dose of $10^{17}$ ions/cm$^3$ [80]. The effect of He irradiation at both high and low energies is imperative to understand the potential for use in advanced nuclear reactors [81,82]. At this high fluence, the HEC was observed to maintain crystallinity after irradiation and from temperatures ranging from 25-1500° C. The peak lattice expansion observed (0.78%) was less than the binary ZrC (0.91%) and the HEC exhibited superior recovery (~0.1% residual expansion after annealing) as compared to ZrC (0.2%). The most significant and promising result of this study was the small He bubble sizes which did not exhibit ripening or percolation in the HEC, leading to a much smaller dislocation density as compared to ZrC. This reduction in He ion irradiation damage, coupled with the heavy and self-ion implantation resilience suggests great potential for use in these extreme environments.

A recent study of $(La_{0.2}Ce_{0.2}Gd_{0.2}Er_{0.2}Tm_{0.2})_2(WO_4)_3$, HEC powder has shown promise in as radiation shielding in a mixed thermal neutron and γ-ray radiation environment [83]. The authors sought to optimize γ-ray shielding in the 40-88 keV range and the thermal neutron shielding though high absorbing cross section of the rare earth metals, in combination with predicted improved radiation tolerance from the high entropy material. A range of HEC powder mass fractions (0%, 10.45%, 10.92% and 31.82%) were added to resin epoxy with surface roughness and particle uniformity measured using electron microscopy and x-ray diffraction analysis. The respective shielding ability of the prepared samples was assessed through neutron radiography and γ dosimetry. This resulted in good γ-ray shielding in the 40-88 keV range, with some reduction in efficiency at higher γ-ray energies. Although the high entropy structure of the HEC powder was selected for its improved radiation damage resistance, the degree of radiation damage was not assessed in this work in post-irradiation analysis. HEC powder-based mixed radiation environment shielding shows potential for continued development, especially if explored with higher fluences for longer duration.

*2.3.2 Actinide Bearing High Entropy Ceramics*

Oxides, particularly pyrochlores, have been explored as potential candidates for immobilization of actinides in the nuclear fuel cycle. This owes largely to the observed resilience of the pyrochlore structure to radiation effects. Whether intended for the disposition of or the future use of actinides, safe and stable storage of actinides is a requirement for nuclear security and environmental responsibility. Radiation tolerance of the matrix holding the actinide element is imperative as material degradation, expansion, or amorphization can all affect the mobility of actinides. The early pyrochlore oxides explored as matrix materials for immobilization of actinides were zirconates ($A_2Zr_2O_7$) and titanates ($A_2Ti_2O_7$). Interestingly, even these early materials took on a "high entropy"-like composition such as the titanate pyrochlore $(Ca,Gd,Pu,U,Hf)_2Ti_2O_7$ [84]. This composition demonstrates the flexibility of the pyrochlore structure and the potential for extension of other simple compositions toward the entropy stabilized and high entropy regimes.

One of the most extreme environment examples is that of the nuclear fuel used in the reactors for nuclear thermal propulsion (NTP) rocket engines for deep space transit. Fuel in these systems is expected to survive temperatures in excess of 2600 K and be in direct contact with high pressure flowing $H_2$ propellant. Early fuel systems in the US Rover/NERVA program were solid solution (ZrU)C [85–88], however, more recent literature [89–91] recommends fuels of three component solid solutions such as (ZrNbU)C and (ZrTaU)C – pushing into today's "medium" entropy classification of HECs. These fuels exhibited higher stability than any of their respective counterparts and allowed for enhancement of the operational temperatures as the presence of Zr and Nb in UC elevates the melting point well beyond 3300 K while also maintaining hydrogen corrosion resistance. Interestingly enough, research has shown that the rock salt structure that UC forms can support not only metal size solutionization, but also nonmetal site solutionization with the formation of carbonitrides (C,N), oxycarbides (O,C) and oxycarbonitrides (O,C,N) [92–95]. The implication for HEC could possibly allow for additional complexity in these ceramic stoichiometries and even higher levels of entropy stabilization. The additional stability (owing to a more favorable energy of formation) should lend further durability and resilience to ionizing radiation as compared to the performance of standard materials.

Two recent reports focusing on high entropy pyrochlores aimed to investigate this hypothesis. The first, utilizing Ce as a stand-in for Pu and Gd as a stand-in for minor actinides, explored the $(Eu_{1-x}Gd_x)_2(Ti_{0.2}Zr_{0.2}Hf_{0.2}Nb_{0.2}Ce_{0.2})_2O_7$ system [96]. This study observed a leaching rate in deionized water which was orders of magnitude lower than that of ternary oxides. The authors attribute this to the increased stability possible with compositional complexity, though further exploration of structural and radiation effects will be vital to exploring this material as a potential candidate for the immobilization of radioactive wastes. In the second report, uranium containing high entropy titanate pyrochlores were explored in a systematic way that increased the degree of complexity in a series of compositions [97]. The study found a strong correlation to changes in thermal conductivity and suggested the role of variance in cation size in determining stability and physical properties of the oxides. This study nicely displays the possibility of U incorporation but fails to exceed 5 atomic percent U in the system, a limitation which has extended past 30% in other studies with compositionally complex pyrochlore oxides. Furthermore, for better understanding of their candidacy for nuclear materials immobilization, these materials require understanding of stability in extremes of radiation, prompting future ion, neutron, β, and γ irradiation studies.

*2.3.3 High Entropy Oxides*

Much of the early oxide work focused on pyrochlores, whose phase stability and flexibility promoted resistance to radiation induced amorphization [84]. Studies of the pyrochlore oxides have perhaps been amongst the most diverse of the high entropy oxides (HEOs), with exploration of their physical properties for thermal barrier coatings [46] and fundamental functional properties related to magnetic structure [98]. However, a vastly expanding area of research is to the radiation tolerance of high entropy pyrochlores. This work is largely informed by the multitude of potential structures discovered by computation methods [26], as well as early results in HEOs, demonstrating structural tunability beyond the parent compound structures [14]. These early studies, particularly with regards to heavy ion irradiation, show limited benefit of high entropy compositions in resistance to amorphization as compared to ternary pyrochlores [57,99], as illustrated in Fig. 4, for $(Yb_{0.2}Tm_{0.2}Lu_{0.2}Ho_{0.2}Er_{0.2})_2Ti_2O_7$ [99] and $Ho_2Ti_2O_7$ [100]. However, some high entropy pyrochlores have shown modest improvement of radiation tolerance attributable to the large lattice distortion of the high entropy ceramic [101]. A study across many multicomponent pyrochlore ceramics showed $(La_{0.2}Nd_{0.2}Sm_{0.2}Eu_{0.2}Gd_{0.2})_2Zr_2O_7$ and $Gd_2(Ti_{0.25}Zr_{0.25}Hf_{0.25}Ce_{0.25})_2O_7$ to hold promise of enhanced radiation resistance under irradiation with 800 keV $Kr^{2+}$ ion beams [102]. This study showed an example expanding the design of the immobilization matrix for high level radioactive waste storage applications.

These results demonstrate that generating by-design radiation tolerance of high entropy pyrochlores is heavily material dependent and demands more focused efforts to explore their potential for applications in extreme environments. First, the role of cation size variance as it correlates to radiation hardness, including where ionic radii constraints are no longer as limiting within a crystal structure, should be explored. In the studies mentioned above, a large variation of cation sizes is explored and in each case is expected to lead to very different local microstructures, bond angles, and defects as was observed in other HEOs [9,33]. These research needs and a general outlook towards high entropy materials in radiation environments is discussed below.

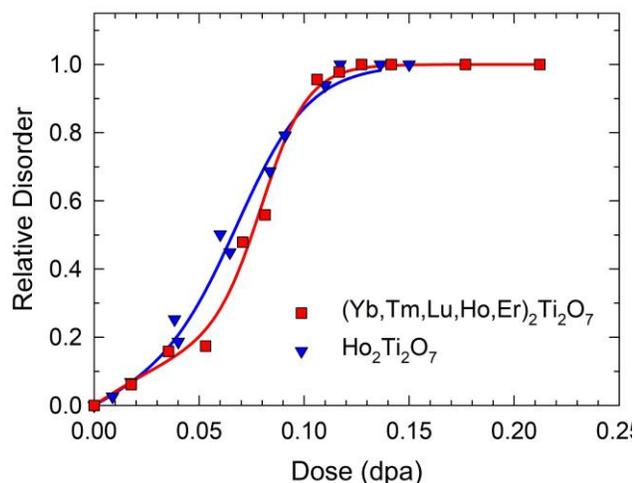

Fig. 4. Relative disorder at the damage peak as a function of local dose in single-crystal $(Yb_{0.2}Tm_{0.2}Lu_{0.2}Ho_{0.2}Er_{0.2})_2Ti_2O_7$ [78] and $Ho_2Ti_2O_7$ [79] irradiated with Au ions at room temperature.

**2.4 Outlook and Research Needs**

The exploration of high entropy materials for applications in extreme environments is still in early development. Carbides have shown enormous potential in high temperature, high radiation environments already, having several studies demonstrating their increased radiation hardness and ability to repair ion-induced radiation. High entropy nitrides and diborides are other refractory systems that may hold promise. Further, systematic studies of these various high entropy compositions will aide our understanding towards what drives these favorable results. Reactor experiments with the intent of exposing HECs to a high neutron and γ fluence environment would allow for study of HECs in more relevant conditions for a fission reactor core, such as Idaho National Laboratory's (INL) Advanced Test Reactor (ATR) and Transient Reactor Test Facility (TREAT), for steady-state or transient radiation environment studies, respectively. Considering the high cost and scheduling difficulty inherent in reactor materials testing, dual and potentially tri-beam studies would also improve representation of the types of radiation present in fission and potential fusion reactors (e.g. neutrons, γ-rays, ions) and develop a more complete view as to the HEC's resilience to such extremes.

The stability of high entropy materials also makes them interesting as potential hosts for radioactive elements across the actinide group. Early work has shown that large amounts of actinides can be incorporated into compositionally complex pyrochlores and that those pyrochlores can exhibit radiation hardness and stability leeching of contamination [84,97]. While this work has been limited, there is a long history of compositionally complex materials as a path towards actinide immobilization, which points to great potential to utilize the added stability and the minimization of the energy of formation that high entropy materials afford through manipulation of complex compositions. Furthermore, the high entropy approach could yield rich results both in synthesis and chemical flexibility of actinide nitrides [103]. These are attractive a nuclear fuels but often have complex reaction pathways and stability issues in reactive environments [104], alluding to the attractiveness of compositional complexity as a way to promote robustness.

HEOs are perhaps the least studied thus far in their resilience to radiation extremes. Pyrochlores thus far have shown mixed results. To better understand what drives improved resilience to radiation in some high entropy oxides, systematic studies relating the local and global structures of these oxides to their amorphization through a series of doses and range of temperatures are vital. Other crystal structures are also emerging, such as spinels [105,106] and high entropy garnets [107]. Here, chemical flexibility combined with the ability for cations to occupy both the tetrahedral and octahedral sites may drive self-repairing behaviors which limits or eliminates progress towards amorphization because of irradiation. Showing some promise, initial studies on $(AlCoCrCu_{0.5}FeNi)_3O_4$ have shown indication of autonomous self-repair processing of damage from electromagnetic irradiation and mechanical impact warranting further investigation across compositions and environments [108]. Other applications, exploring the absorption of thermal neutron and gamma-rays by HECs show modest promise as radiation shielding materials [83], are still emerging as this young field grows.

### 3. Functional High Entropy Ceramics in Extreme Environments

**3.1 A New Path Towards Designing Materials**

A great deal of excitement has surrounded functional high entropy materials – extending from the largely structural focus of high entropy alloys. The root of this excitement is the addition of an anion sublattice which can and does lead to strongly correlated and nontrivial behaviors – from magnetism to ultralow thermal conductivity to scintillation. Due to the general ability to tune compositions, owning to the added stability of configurational entropy, these behaviors have been observed to be continuously controllable where critical temperatures and phases can be predicted and manipulated to a given requirement [19]. Beyond tuning properties, high entropy materials are beginning to demonstrate functional behaviors outside of known phase diagrams. These include a broad bandwidth of phenomena from monolithic exchange bias [109] to high ion conductivity [23]. This control emerges from local phase frustration driven by compositional complexity, represented in Fig. 5 where the site-to-site cation disorder, represented by the open circles of varying size, leads to different interactions and competing (underlying blue, red) phases. To focus the review, we broadly cover magnetic, electronic, and optical properties which benefit from the tunability and stability of high entropy materials.

## 3.2 Magnetic materials

Much of the most promising work in exploring magnetism in high entropy materials has been in the subfield of high entropy oxides [12,14,110,111]. Unlike trivial itinerant magnetism through Ruderman-Kittel-Kasuya-Yosida (RKKY) interactions [112,113] in high entropy alloys, the presence of the anion sublattice enforces electronic correlation which gives rise to the unusual and exciting electronic and magnetic properties observed in more traditional strongly correlated materials. Functional magnetic HEOs come in many different varieties including spinel [105,106,114], perovskite [34,42,115,116], Ruddlesden-Popper [41,117], pyrochlore [26,118,119], and rocksalt [120,121] structures. The continuously tunable nature of HEOs – allowing for careful, uniform doping across broad and otherwise unreachable areas of phase space – has already shown promise beyond traditional materials. As an example, continuously tunable long range magnetic order to phase frustration leading to monolithic exchange bias [19,109,122], strain tunable spin textures [123], and decoupling of critical order parameters not present in traditional analogues [33] have all been observed in HEOs.

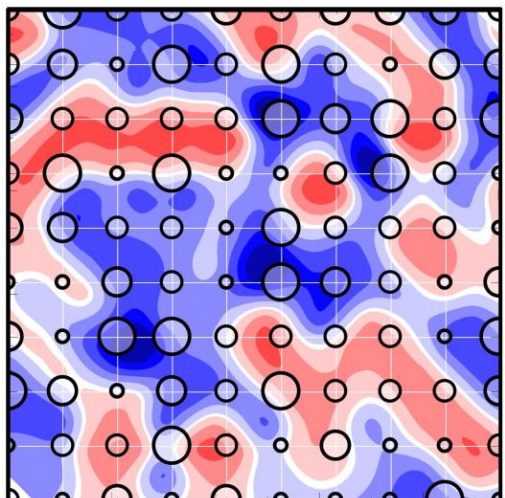

Fig. 5. A representation how phase frustration (phase 1 in red and phase 2 in blue) can be driven by compositional complexity (atoms represented by open black circles where the size can represent cation radius, spin, charge, etc.).

The tunability of magnetic phase and phase degeneracy in magnetic HEOs is one of the most promising in tailoring them to device applications. This tunability, and the ability to predict magnetic phase, is well established in the high entropy transition metal oxide La(Cr$_{0.2}$Mn$_{0.2}$Fe$_{0.2}$Co$_{0.2}$Ni$_{0.2}$)O$_3$ (L5BO) [19,124]. Here it was shown that as a function of Mn partial fraction the dominant magnetic phase could be tuned from antiferromagnetic to ferromagnetic with monolithic exchange bias occurring when the two phases coexist. This tunability seems to be somewhat universal in HEOs having been established in spinels [5] and rocksalts [125]. Overall, it appears the main versatility of magnetic HEOs is their tunability and designer phase degeneracy, though their frustrated nature and ability to stabilize phases outside of established phase diagrams is sure to yield more exotic magnetic states such as chiral spin textures and controllable magnetization dynamics.

## 3.3 Electronic materials

Research into HECs as electronic materials has been much broader than that of magnetic materials. This in part is due to the broadness of the field, randing from packaging materials to sensing to microelectronics. The relevant HECs span carbides [126], silicides [127], oxides [33,128,129], chalcogenides [130,131] and beyond [132]. Each of these arenas have potential for devices operating in extremes, i.e., carbides at high temperatures [47] and oxides in high pressure, hydrogen rich, and high radiation environments [57,99]. As an example in the HEOs, the functional properties offer a wide diversity of behaviors and include superconductivity [129,133], tunable metal to insulator transitions [33,134], bandgap tunability [135,136], and ferroelectricity [137]. Additionally, it is worth considering that many traditional microelectronic devices are generated in heterostructured films and 2D materials, where interfacial effects derived from surface decoration [138] and gating [139,140] can be used to set properties in the functionally active bulk layer. The inclusion of high entropy materials into these multilayer device architectures is entirely unexplored but offers many tantalizing possibilities to bridge robustness against harsh environments to on-chip applications.

An area showing immediate promise for applications in the HEOs are dielectric and ferroelectric materials. The colossal dielectric constants [24] observed appear to be somewhat universal – driven by charge compensation mechanisms necessary to stabilize compositional complexity. Many mechanisms might drive these electrical properties, from the atom-level domain sizes to electronic phase frustration. Much of the same nano-scale phenomena can drive promising ferroelectric responses in HEOs [21,137]. The current paradigm in designing materials with ferroelectric responses is very much in agreement with the HECs community – aiming towards substitutional doping and local compositional complexity to drive electromechanical responses, tunable dielectric responses, and polar domain formation. Overall, these frustration-driven electronic phenomena are a highly promising area of future study in the HECs field.

Another class of materials, high entropy chalcogenides [131,141–143], may also provide an interesting pathway towards control of functionality with resilience to extremes. Quantum materials research has focused heavily on topology in materials [144], with a great deal of focus on layered ultraclean chalcogenides. As an example, $MnBi_2Te_4$ can exhibit the quantum anomalous Hall effect [145] subject to its highly sensitive to small degrees of doping and stacking order [146]. However, the resistance of materials such as these to even exposure to oxygen can destroy or alter the properties they exhibit [147]. In fact, superconductivity has been observed in high entropy tellurides with unconventional robustness to extreme pressures [142] – a strong indicator towards the functional stability of chalcogenides synthesized by this approach. Exploring the ability to host these exotic states in high entropy equivalents and how the added stability affects their resilience may yield a pathway towards realizable next-generation electronics.

**3.4 Optical Properties**

The study of optical properties in HECs extends from understanding band structure [136] to scintillation properties [148]. As in the case of other functional properties the broadly tunable compositions, thermal, and structural stability lend to the ability to engineer properties to a certain application. An example of this tunability was recently demonstrated in high entropy titanate perovskite oxides [149]. Here the authors used a systematic approach to tune the density of states by varying the composition of the perovskite A-site and showed as a result they could tune the reflectance, band gap and thermal production. While this result was promising, it did bring into question the ability to control oxygen stoichiometry in these oxides, which is known to be a driver of optical properties in a wide range of oxides. In particular because of a highly tunable oxygen vacancy density in spinel HEOs, it has been shown that the reflectance can be heavily suppressed for a single cation composition [150].

The study of luminescent HEOs, while limited, has shown promise towards applications in scintillation and laser technologies. The potential of rare earth aluminates, which have high optical quality and attractive luminescence properties, are an area where the stability of high entropy materials can cater towards eventual applications. Early work on single crystal high-entropy aluminates have been shown to extend the temperature range of structural stability in $RE_4Al_2O_9$ compounds [151]. This could expand into other structures, such as garnets [107], where mixing multiple cation species can cause improvement in scintillation properties through further incorporation of luminescent ions. This has already been seen by incorporation of cerium and gadolinium (both active luminescent centers) in perovskite thin films [148], where clear photoluminescence was observed. These early results show promise, particularly in enhancing performance at high temperatures [151], but further work must be done to understand the role of the configurational entropy in the observed properties.

Band engineering and absorbance control are additional areas in which high entropy and compositional complexity seem to lend superior tunability and stability as compared to traditional materials. Lattice distortions are well known to control band tuning, having been demonstrated in oxide semiconductors by continuous tuning via He ion implantation [152]. As discussed above, the high entropy synthesis method has similar control over structural flexibility of the lattice, both in type and in dimension. This caters towards control over local distortions, oxygen content, cation size and type all of which can aide in band gap engineering. Fluorite HEOs have shown this type of tunability in the bandgap with controllable changes up to 1.8eV by modifying oxygen stoichiometry [8,136], though a great deal of compositions and structures are yet to be explored. As a final note, combined with bandgap control, the optical absorbance of high entropy borides also show great promise – with oxidation and degradation resistance at extreme temperatures [20]. These borides exhibit low thermal emittance and some (potentially tunable) energy selectivity making them strong candidates for high-temperature solar applications.

**3.5 Outlook and research needs**

HECs rely on having many different elements hosted on a crystalline lattice to protect themselves against radiation effects. Atoms are already randomly distributed on the lattice and their functionality depends on this randomness of atomic arrangement. In theory, knocking elements out of position has a greatly reduced effect on functionality since disorder is an inherent factor in producing magnetic and electronic properties. To test this theory, there is a distinct need for increased experimental and theoretical work to understand how these properties evolve as a function of external stimuli such as irradiation, chemical, and thermal extremes.

Theoretical studies of high entropy materials have been largely oriented towards structural materials like high entropy carbides and alloys [13,15,153]. However, their study does not extend considerably past phase stability, with a relatively small number of papers discussing magnetic [19,121] and electronic properties [126]. However, there is still a need to extend these studies to understand phase stability in extremes and the impacts of formation of different defect types and densities on functional properties. While the simple understanding of the Gibbs free energy of formation can lend to our understanding of stabilization of high entropy materials, this simplification needs to be extended to model real systems and guide our exploration of future functional high entropy materials for use in extremes.

Experimental results have been more illuminating, particularly in radiation resistant materials. Two classes of materials, spinels and pyrochlores, are the main subclasses of HEOs which are being explored. Pyrochlore HEOs are some of the only high entropy systems which have been stabilized in bulk single crystals [98], providing a particularly promising platform from which to study the effects of different types of radiation (i.e. high energy neutron, γ, etc). While there have been some promising results [57], there is a great deal to be done to understand the radiation tolerance of these complex crystal structures. As example, a spinel structure has a great propensity towards site inversion. To understand if the material evolves with radiation dose, particularly with regards to vacancy formation and recombination, a careful understanding of very local disorder should be explored. X-ray absorption spectroscopy (XAS) and extended x-ray absorption fine structure (EXAFS) measurements are both particularly useful in monitoring changes at the very local length scale and with element specificity [6,12,41]. Studies with an understanding of local disorder and microstructure [5] are necessary to understand site inversion and other defects formed during exposure to ionizing radiation and other types of extremes.

4. **Conclusion**

The emergence of compositional complexity as a pathway towards new material design through the so called "high entropy" approach lends a great deal of promise in applications for extreme environments. While the field is young, HECs, owning to their added stability, have already shown enhanced performance in high temperature, chemically harsh, and high radiation environments. Focusing on high fluence radiation environments these materials have begun to show a tolerance beyond their simpler counterparts, particularly due to the resistance to creating dislocation loops leading to material failure. Beyond their structural stability, these materials have demonstrated broad tunability of functionality from magnetism to optical properties which show a similar robust nature – showing promise in applications such as microelectronics for extreme environments. This same diversity of functional properties makes HECs particularly interesting for barrier coatings of various types. These extend from corrosion-resistant coatings to high temperature, low thermal transport barrier coatings where the current state of the art is being challenged by development of new HECs. The growth of the HECs field has been exponential since their emergence in 2015 and the ability to push beyond known phase spaces to develop new materials with a variety of structural, mechanical, and functional properties in this way is sure to revolutionize and cater to their application in extreme environments.

## **Acknowledgements**

This work was supported by the U.S. Department of Energy (DOE), Office of Science, Basic Energy Sciences (BES), Materials Sciences and Engineering Division and the NNSA's Laboratory Directed Research and Development Program at Los Alamos National Laboratory. Los Alamos National Laboratory, an affirmative action equal opportunity employer, is managed by Triad National Security, LLC for the U.S. Department of Energy's NNSA, under contract 89233218CNA000001. The contribution of W.J. Weber was supported by the National Science Foundation under Grant No. DMR-2104228.